\documentclass{article}
\usepackage{spconf,amsmath,graphicx}
\usepackage{amssymb} % checkmark
\usepackage{cite}
\usepackage{url}
\usepackage[colorlinks,
linkcolor=green, 
anchorcolor=blue,
citecolor=blue]{hyperref}

\usepackage{enumitem}
\setlist{nosep, leftmargin=14pt}

\usepackage{mwe} % to get dummy images

\title{SAMIHS: Adaptation of Segment Anything Model for Intracranial Hemorrhage Segmentation}
\name{Yinuo Wang, Kai Chen, Weimin Yuan, Cai Meng, XiangZhi Bai %, \IEEEmembership{Member, IEEE}
\thanks{Co-first authors: Yinuo Wang, Kai Chen.}}
\address{Beihang University}

\begin{document}
%\ninept
%
\maketitle
\begin{abstract}
Segment Anything Model (SAM), a vision foundation model trained on large-scale annotations, has recently continued raising awareness within medical image segmentation.
Despite the impressive capabilities of SAM on natural scenes, it struggles with performance decline when confronted with medical images, especially those involving blurry boundaries and highly irregular regions of low contrast.
In this paper, a SAM-based parameter-efficient fine-tuning method, called SAMIHS, is proposed for intracranial hemorrhage segmentation, which is a crucial and challenging step in stroke diagnosis and surgical planning.
Distinguished from previous SAM and SAM-based methods, SAMIHS incorporates parameter-refactoring adapters into SAM's image encoder and considers the efficient and flexible utilization of adapters' parameters.
Additionally, we employ a combo loss that combines binary cross-entropy loss and boundary-sensitive loss to enhance SAMIHS's ability to recognize the boundary regions.
Our experimental results on two public datasets demonstrate the effectiveness of our proposed method. Code is available at \url{https://github.com/mileswyn/SAMIHS}.
\end{abstract}
\section{Introduction}
\label{sec:intro}

Intracranial Hemorrhage Segmentation (IHS), a significant and challenging task in medical image analysis, serves as a prerequisite for the diagnosis and surgical planning of hemorrhagic stroke \cite{li2020deep}. 
Due to the large scale of hemorrhage attribute variability in each scan, the inherent low contrast, and the complex interaction between hemorrhage and normal regions, it is difficult to delineate the hemorrhage even for experienced doctors. 
In recent years, large numbers of CNN-based and transformer-based image segmentation models have emerged with the wave of deep learning. However, the design of these model structures still relies on experts' experience, which makes the performance fluctuate greatly when encountering various tasks. 
%(task-specific)

Recently, the Segment Anything Model (SAM) \cite{kirillov2023segment}, a vision foundation model trained with over 1 billion masks, has exhibited its remarkable zero-shot segmentation capability across diverse vision tasks with user prompts, including points, bounding boxes, and texts.
Despite its subpar performance on many medical image segmentation tasks, SAM simultaneously showcases its potential through fine-tuning techniques \cite{mazurowski2023segment}.
Instead of directly fine-tuning a large number of parameters of the whole SAM, an alternative flexible way is to fine-tune a portion of SAM's parameter or utilize parameter-efficient fine-tuning (PEFT) techniques \cite{houlsby2019parameter,hu2021lora,jia2022visual,lian2022scaling,dong2023efficient}, in order to transfer the pre-trained SAM to a specific medical image segmentation task.

In this paper, we propose SAMIHS, a novel method that integrates parameter-refactoring adapters into SAM for efficiently adapting SAM to IHS task.
Compared to previous methods \cite{zhang2023customized,lin2023samus,wu2023medical}, SAMIHS consolidates the correlation between adapters and the knowledge extracted from SAM by establishing two independent up/down linear projections of low-level features at each layer.
In addition, a boundary-sensitive objective function is exploited to improve the perception of the low-contrast hemorrhage regions, and it is designed as a task-specific loss in our fine-tuning strategy.
We conducted experiments on two publicly available CT datasets to demonstrate the effectiveness and superiority of SAMIHS.
The contribution of this paper can be summarized as follows:
\begin{itemize}
    \item To handle the IHS task, we designed SAMIHS, a task-specific model extended from SAM, which completed a puzzle piece of SAM's application in the medical imaging domain.
    \item We adopted novel parameter-refactoring adapters and a boundary-sensitive loss in our fine-tuning strategy. These components effectively improve the model's performance.
    \item We evaluated our proposed SAMIHS on two publicly available datasets, and compared the results with some CNN-based, transformer-based, and PEFT-based state-of-the-art methods.
\end{itemize}

\begin{figure*}[!t]
\centerline{\includegraphics[width=2\columnwidth]{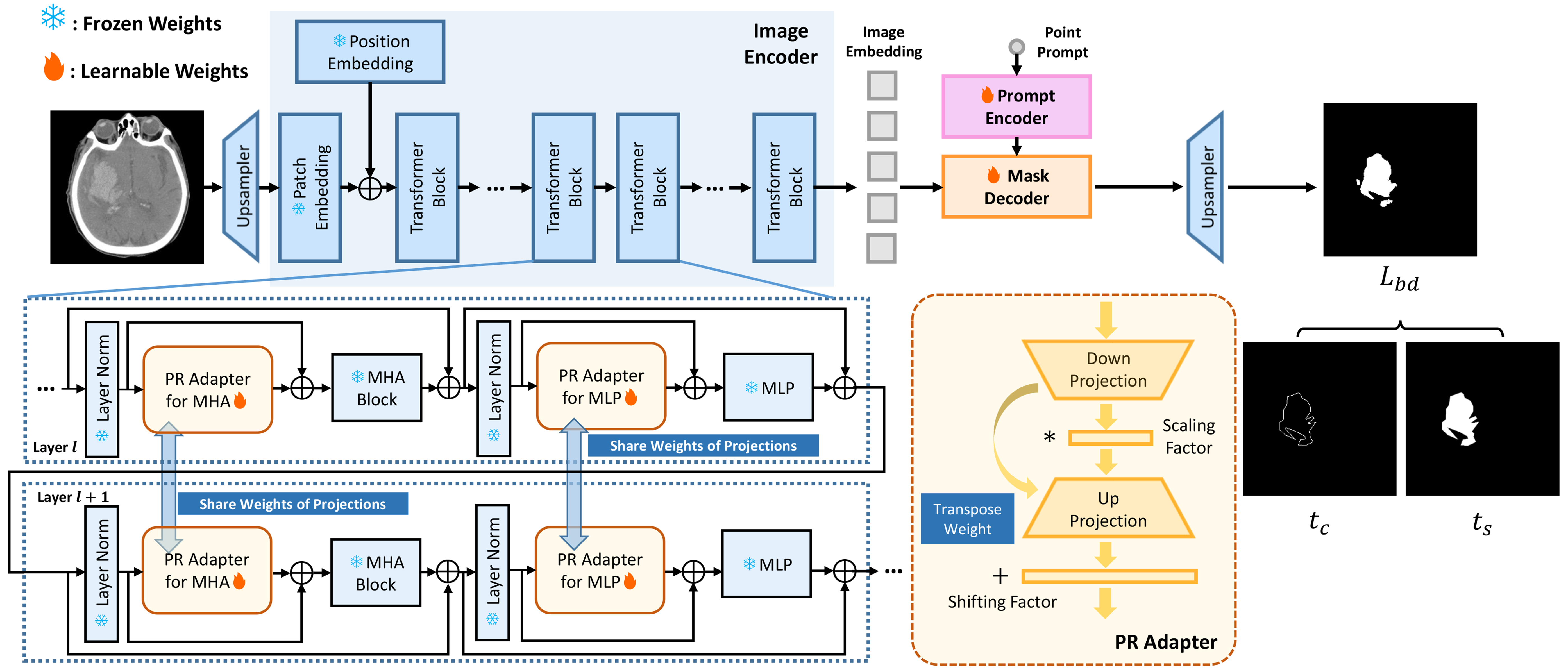}}
\caption{The overall architecture of our proposed SAMIHS.}
\label{fig1}
\end{figure*}

\section{Related Works}
\label{sec:format}

Although SAM has shown impressive results in a broad range of natural scenes, it has been proven to lack equal advantages when directly applied in medical domains \cite{mazurowski2023segment}. 
Based on this, MedSAM \cite{ma2023segment} froze the massive image encoder and the prompt encoder of SAM, and fine-tuned the lightweight mask decoder on generous medical images.
In previous studies, visual tuning methods have been considered to be effective in adapting the foundation model to specific downstream tasks.
Especially, adapter tuning incorporates different forms of adapters into multiple locations of the foundation model, resulting in significant performance enhancement with less additional parameter cost \cite{houlsby2019parameter,hu2021lora,lian2022scaling,jia2022visual,dong2023efficient}. 
Some work has introduced this approach to the fine-tuning of SAM.
For example, SAMed \cite{zhang2023customized} applied a low-rank-based strategy (LoRA) in the frozen image encoder, fine-tuning the LoRA layers, prompt encoder and mask decoder together on the Synapse multi-organ dataset.
MSA \cite{wu2023medical} adopted down-RELU-up adapters in each transformer block of the image encoder. The adapters were incorporated into the attention layers and MLP layers in serial and parallel manners respectively.
SAMUS \cite{lin2023samus} designed a learnable CNN branch parallel to SAM's image encoder while adding feature adapters for transformer blocks and position adapter for positional embedding.
Our proposed SAMIHS optimized the adapter design and fine-tuning loss function while also conducting specialized evaluations on intracranial hemorrhage segmentation.

\section{Method}
\label{sec:pagestyle}

\subsection{Overview}

As depicted in Fig.~\ref{fig1}, the overall structure of SAMIHS inherits from SAM.
We add parameter-refactoring adapters (PR adapters) in the image encoder while maintaining the original architecture of the prompt encoder and mask decoder.
Given a head CT scan $x \in \mathcal{R}^{H \times W \times C}$ as input, SAMIHS first upsamples it to $x_{up} \in \mathcal{R}^{2H \times 2W \times C}$ and passes it forward through the image encoder. 
Then the mask decoder fuses the image embedding and the prompt feature to predict the low-resolution mask $ p_{low} \in \mathcal{R}^{H/2 \times W/2}$.
Through another upsampler, the final prediction map $ p \in \mathcal{R}^{H \times W}$ can be obtained.

In SAM, the input of the prompt encoder can be sparse (points, boxes, and text) or dense (masks).
Considering the convenience of interactive segmentation on medical images, SAMIHS explores the simplest single-point prompt in this paper.
During training, we freeze the original image encoder and fine-tune the PR adapters, prompt encoder, and mask decoder of SAMIHS.

\begin{figure}[!t]
\centerline{\includegraphics[width=1.1\columnwidth]{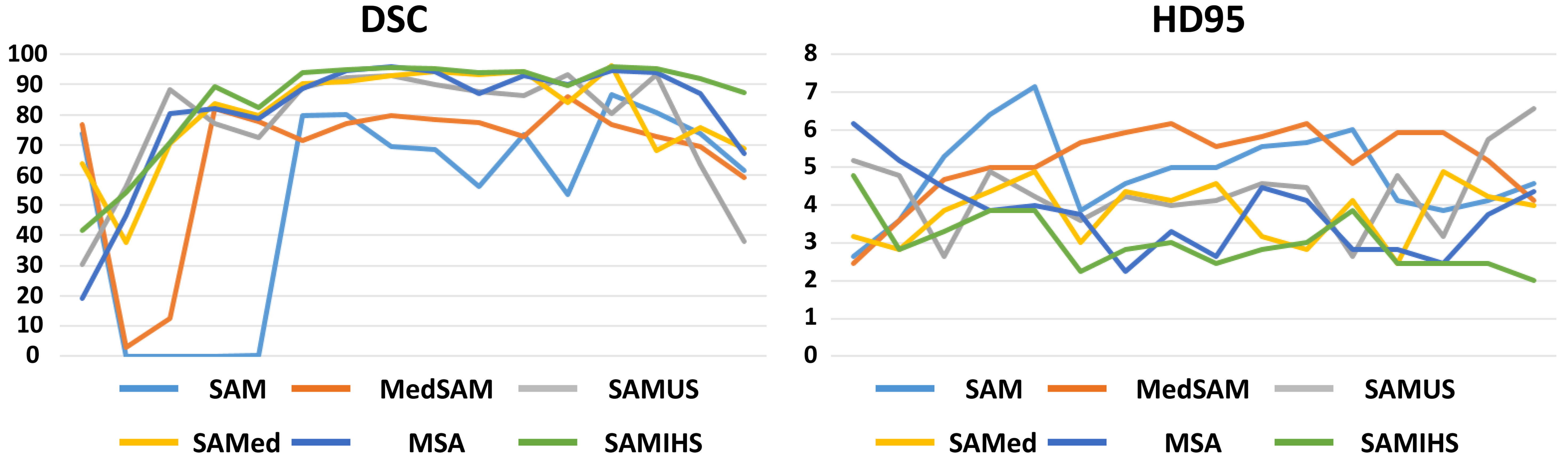}}
\caption{The line charts of slice by slice prediction results in a single CT case.}
\label{fig2}
\end{figure}

\subsection{Parameter-Refactoring Adapter}

Previous studies have introduced separate learnable parameters to different layers of SAM in various forms.
Inspired by these works, we explore a novel approach to configure a unified linear mapping setting for adapters at the same positions in different transformer blocks, in order to enhance the efficiency of adapters.
Specifically, as shown in Fig.~\ref{fig1}, given an input feature $m_{\text in} \in \mathcal{R}^{h \times w \times c}$, it successively passes through the symmetric down-projection $W_{\text down} \in \mathcal{R}^{c \times c^{\prime}}$ and up-projection $W_{\text up} \in \mathcal{R}^{c^{\prime} \times c}$.
The forward process in PR adapters can be formulated as follows:
\begin{equation}
\begin{aligned}
m_{\text out} &= m_{\text in} W_{\text down} R_{l} W_{\text up} + B_{l} + m_{\text in}
\end{aligned}
\end{equation}
where $R_{l} \in \mathcal{R}^{c^{\prime}}$ and $B_{l} \in \mathcal{R}^{c}$ represents the scaling and shifting factors respectively. 
The scaling factor $R_{l}$ and shifting factor $B_{l}$ are independent for each adapter in each layer, aiming to dynamically refactor multi-layer features to promote adaptation performance.
Since SAM adopts windowed attention that partitions the features before MHA blocks, we configure two PR adapters before MHA blocks and MLP by default to facilitate flexibility in adjustment within each transformer block.
Thus, the modified transformer block can be formally described as follows:
\begin{equation}
\begin{aligned}
m^{(l)\prime} &= \text{MHA}(\text{PR}_{\text{MHA}}(\text{LN}(m^{(l-1)}))) + m^{(l-1)}
\end{aligned}
\end{equation}
\begin{equation}
\begin{aligned}
m^{(l)} &= \text{MLP}(\text{PR}_{\text{MLP}}(\text{LN}(m^{(l)\prime}))) + m^{(l)\prime}
\end{aligned}
\end{equation}
note that $\text{PR}_{\text{MHA}}$ and $\text{PR}_{\text{MLP}}$ are independent while their weights across different transformer blocks are shared to improve the efficiency of PR adapters.

\subsection{Loss Function}

Unlike the learning objects of SAM in natural scenes, the intracranial hemorrhage has low contrast with surrounding normal tissue, leading to blurry boundaries of hemorrhage, which is prone to segmentation errors. Inspired by relevant research \cite{karimi2019reducing,kervadec2019boundary,sun2023boundary}, we introduce a boundary-sensitive loss function in training SAMISH and combine it with binary cross entropy loss to improve the performance of hemorrhage prediction.
Given a pair of prediction map $P$ and ground truth G containing hemorrhage target, our goal is to decrease the difference area $P \cup G - P \cap G$ and increase the intersection $P \cap G$, especially to increase the intersection of boundary regions.
To this end, the boundary-sensitive loss can be written as follows:
\begin{equation}
\begin{aligned}
{{L}_{bd}} &= 1 - \frac{2\gamma TP}{2\gamma TP+FP+FN}
\end{aligned}
\end{equation}
among which ${\gamma}$ is a dynamic factor that regulates the sensitivity of boundary perception,
\begin{equation}
\begin{aligned}
{\gamma} &= 1 - \frac{{t}_{c}}{{t}_{s}}
\end{aligned}
\end{equation}
where ${t}_{c}$ and ${t}_{s}$ respectively denote the target's boundary length and the target's size. 
The overall loss of SAMIHS is as follows:
\begin{equation}
\begin{aligned}
{L} &= \lambda_{1}{L}_{bd} + \lambda_{2}{L}_{ce}
\end{aligned}
\end{equation}
where the trade-off parameters $\lambda_{1} = \lambda_{2} = 0.5$.

\begin{table}[t]
\caption{Quantitative comparison of different methods in Dice and HD95 using five-fold cross validation. The "Params" in the table indicates learnable parameters.}
\label{table1}
\centering
\resizebox{0.45\textwidth}{!}{
\begin{tabular}{c|cc|cc|c}
\hline
Dataset         & \multicolumn{2}{c|}{BCIHM}  & \multicolumn{2}{c|}{Instance}  &             \\ \hline
Method          & Dice & HD95  & Dice & HD95 & Params(M) \\ \hline
U-Net           & 50.06    & 3.99     & 62.07    & 4.26     & 7.77               \\
Att-UNet        & 54.29    & 3.89     & 29.74    & 7.14     & 34.88            \\
U-Net++         & 53.80    & 3.92     & 53.67    & 4.86     & 9.16            \\
TransUNet       & 46.47    & 4.05     & 58.23    & 4.44     & 106.17            \\
TransFuse       & 52.14    & 3.86     & 24.83    & 7.63     & 26.57            \\
H2Former        & 48.79    & 4.03     & 31.12    & 5.61     & 33.86            \\
SAM             & 49.32    & 4.29     & 61.46    & 5.04     & N/A            \\
MedSAM          & 51.38    & 4.51     & 51.38    & 4.51     & N/A            \\
SAMed           & 66.13    & 3.56     & 74.99    & 3.77     & 3.93            \\
SAMUS           & 60.29    & 3.85     & 43.85    & 5.46     & 42.60            \\
MSA             & 67.08    & 3.53     & 72.65    & 3.98     & 11.17            \\
SAMIHS          & \pmb{69.77}    & \pmb{3.31}     & \pmb{76.52}    & \pmb{3.71}     & 4.24             \\ \hline
\end{tabular}}
\end{table}

\section{Experiments}
\label{sec:typestyle}

\subsection{Experimental setup}

\textbf{Datasets.} 
We evaluated SAMIHS on two public datasets. 
The two datasets named BCIHM and INSTANCE collected by Hssayeni \cite{hssayeni2020intracranial} and Li et al. \cite{li2023state}, include 36 and 200 non-contrast CT volumes for individuals diagnosed with intracranial hemorrhage with the following types: Epidural, Subdural, Intraventricular, Intraparenchymal, and Subarachnoid. 
In this paper, we used all cases of BCIHM and 100 cases of INSTANCE, which were marked as training set in a MICCAI 2022 Challenge.
Each CT we used in the two datasets has the same original size of 512$\times$512 in the transverse section and 5mm thickness on the z-axis.
To make convincing comparisons, we extracted and shuffled slices in each dataset, on which 5-fold cross validations were conducted for both contrast and ablation experiments.

\noindent \textbf{Preprocessing.}
For all slices, we clipped the intensity by 0.5 and 99.5 percentiles of foreground and then performed Max-Min normalization.
During training, random shift, rotation, and adding Gaussian noise were used as data augmentations.

\noindent \textbf{Implementation Details.}
We utilized SAM's pre-trained ViT-B variant and fine-tuned SAMIHS in 200 epochs.
The training batch size was 2, and the ADAM optimizer was used with an initial learning rate of $5.0\times10^{-4}$. 
The experiments were conducted using Python 3.10 and Pytorch 1.13.0 on an NVIDIA RTX 3090 GPU. 
The Dice score and $95\%$ Hausdorff distance (HD95) were used for evaluation.
\begin{figure}[!t]
\centerline{\includegraphics[width=\columnwidth]{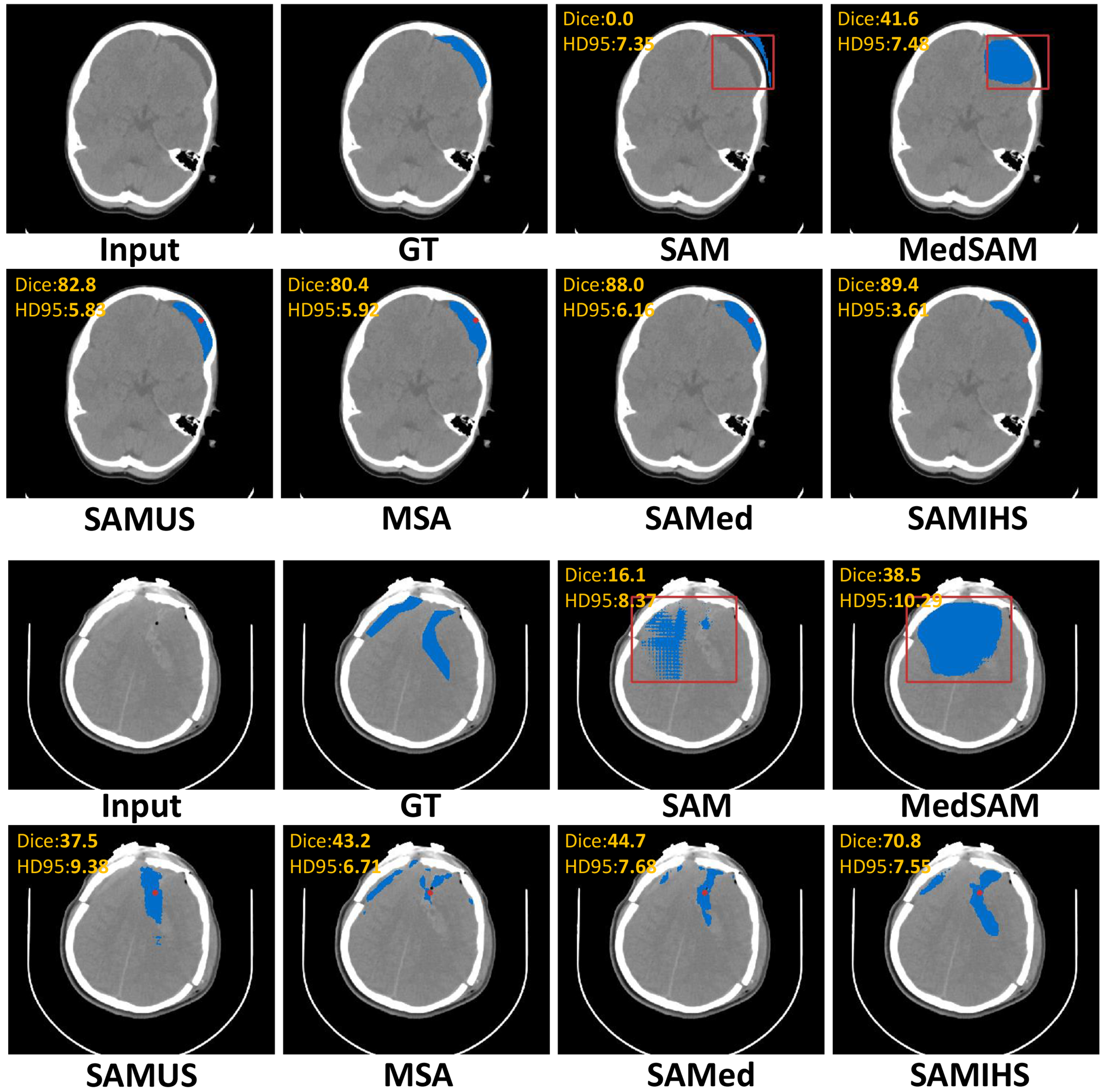}}
\caption{Visual comparison of hemorrhage segmentation results produced by our proposed SAMIHS and other SAM-based methods on the BCIHM dataset. For different methods, we follow their settings and utilize boxes or points as prompts.}
\label{fig3}
\end{figure}

\subsection{Results}

\textbf{Quantitative results.} 
We compared our method with state-of-the-art methods, including three CNN-based models (U-Net \cite{ronneberger2015u}, Att-UNet \cite{oktay2018attention}, and U-Net++ \cite{zhou2018unet++}), three transformer-related methods (TransUNet \cite{chen2021transunet}, TransFuse \cite{zhang2021transfuse}, and H2Former \cite{he2023h2former}), SAM \cite{kirillov2023segment}, and four SAM-based methods (MedSAM \cite{ma2023segment}, SAMed \cite{zhang2023customized}, SAMUS \cite{lin2023samus}, and MSA \cite{wu2023medical}). 
According to Table~\ref{table1}, compared to CNN-based and transformer-related methods, the SAM-based methods, especially our proposed SAMIHS, achieve a significant lead in Dice and HD95.
These advantages are attributed to SAM's knowledgeable pre-trained model and extra visual prompts.
Meanwhile, SAMIHS outperforms other SAM-based PEFT methods (SAMed, SAMUS, MSA) in Dice and HD95 on two datasets at the second least learnable parameter cost.
In addition, to analyze the segmentation performance on 3D volumes, we analyzed the prediction results of several methods on a single CT case slice by slice on the z-axis, as shown in Fig~\ref{fig2}.
SAMIHS demonstrates better accuracy and continuity in predicting hemorrhage with inconsistent shapes.

\noindent \textbf{Qualitative results.}
Fig~\ref{fig3} demonstrates the qualitative results of SAM and several SAM-based methods in the BCIHM dataset.
The prediction of two hemorrhage types, epidural and intraparenchymal, are compared in the upper and lower two rows.
Due to the absence of fine-tuning, SAM and MedSAM are prone to confusion between hemorrhage and surrounding tissues, despite using more precise box prompts.
Among the methods that use single-point as a prompt, SAMIHS shows the best predictive ability.
Particularly, within the context of well-defined subdural hemorrhage predictions, SAMIHS exhibits the highest precision as shown through comparative analysis.
Conversely, when confronted with predictions of intraparenchymal hemorrhage characterized by less distinct boundaries, other methods experience significant performance degradation.
Nevertheless, SAMIHS continues to predict areas and contours of relative precision.
These comparisons highlight SAMIHS's commendable generalization ability when dealing with variations in the shape and contrast of hemorrhagic regions.

\noindent \textbf{Ablation study.}
Our research involved analyzing different variations of SAMIHS in order to evaluate their effectiveness.
The results, as presented in Table~\ref{table2}, indicate that the uses of PR adapters on both MHA and MLP, along with the boundary-sensitive loss, are highly effective in promoting prediction performance.
Note that the performance only marginally decreases when one of the PR adapters on MHA or MLP is removed, but it exhibits a significant decline when both PR adapters are removed simultaneously.

\begin{table}[h!]
\caption{Ablation study on PR adapters for MHA, PR adapters for MLP and boundary-sensitive loss.}
\centering
\label{table2}
\resizebox{0.40\textwidth}{!}{
\begin{tabular}{ccc|cl|cl}
% \toprule[1.0pt]
\hline
\multicolumn{3}{c|}{Method} & \multicolumn{2}{c|}{BCIHM} & \multicolumn{2}{c}{Instance} \\ \cline{1-7} 
 MHA    &  MLP    & BD     & Dice        & HD95     & Dice        & HD95     \\ \hline
    &   &                       & $62.76$        & $3.71$       & $57.49$        & $4.84$                 \\ \hline
&   &\checkmark                 & $63.03$        & $3.66$       & $57.80$        & $4.72$                 \\ \hline
& \checkmark   &                & $68.59$   & $3.39$   & $75.28$   & $3.80$             \\ \hline
\checkmark &  &                 & $68.17$   & $3.56$   & $74.83$   & $3.79$             \\ \hline
& \checkmark  &\checkmark       & $68.98$   & $3.38$   & $75.60$   & $3.79$             \\ \hline
\checkmark    &   & \checkmark  & $68.37$   & $3.45$   & $75.26$   & $3.78$             \\ \hline
\checkmark    & \checkmark  &   & $69.46$   & $3.46$   & $76.09$   & $3.77$             \\ \hline
\checkmark    & \checkmark  & \checkmark  & \pmb{$69.77$}  & \pmb{$3.31$}   & \pmb{$76.52$} & \pmb{$3.71$}  \\ \hline        %   \\ \bottomrule[1.0pt]
\end{tabular}}
\end{table}

\section{Conclusion}
\label{sec:majhead}

Our paper introduces SAMIHS, which is a SAM-based parameter-efficient fine-tuning method designed to improve intracranial hemorrhage segmentation (IHS).
SAMIHS inserts parameter-refactoring adapters into the image encoder, improving the reusability of adapter parameters in adjacent transformer blocks, and adopts a novel boundary-sensitive loss to enhance hemorrhage prediction performance.
Comparative and ablation experiments conducted on two public datasets have demonstrated that SAMIHS outperforms multiple state-of-the-art methods, and all the proposed modules play critical roles in the method.
In future work, we plan to explore extending SAMIHS to an efficient 3D segmentation method and optimizing the current prompt to maximize the potential of our method in IHS task.

\section{Acknowledgments}
\label{sec:print}

This work was supported in part by the National Natural Science Foundation of China under Grants (92148206, 62271016).

\section{Compliance with ethical standards}
\label{sec:ethics}

This research used two public non-contrast head CT datasets for experiments. The first was collected by Hssayeni \cite{hssayeni2020intracranial} and the second was from the INSTANCE challenge \cite{li2023state}. Ethical approval was not required as confirmed by the license attached with the open access data.

% ------------------------------------------------------------------------- 
\bibliographystyle{IEEEbib}
\bibliography{refs}

\begin{thebibliography}{10}

\bibitem{li2020deep}
L.~Li et~al.,
\newblock ``Deep learning for hemorrhagic lesion detection and segmentation on brain ct images,''
\newblock {\em IEEE journal of Biomedical and Health Informatics}, vol. 25, no. 5, pp. 1646--1659, 2020.

\bibitem{kirillov2023segment}
A.~Kirillov et~al.,
\newblock ``Segment anything,''
\newblock {\em arXiv preprint arXiv:2304.02643}, 2023.

\bibitem{mazurowski2023segment}
M.A. Mazurowski, H.~Dong, H.~Gu, J.~Yang, N.~Konz, and Y.~Zhang,
\newblock ``Segment anything model for medical image analysis: an experimental study,''
\newblock {\em Medical Image Analysis}, vol. 89, pp. 102918, 2023.

\bibitem{houlsby2019parameter}
N.~Houlsby et~al.,
\newblock ``Parameter-efficient transfer learning for nlp,''
\newblock in {\em International Conference on Machine Learning}. PMLR, 2019, pp. 2790--2799.

\bibitem{hu2021lora}
E.J.~Hu et~al.,
\newblock ``Lora: Low-rank adaptation of large language models,''
\newblock {\em arXiv preprint arXiv:2106.09685}, 2021.

\bibitem{jia2022visual}
M.~Jia et~al.,
\newblock ``Visual prompt tuning,''
\newblock in {\em European Conference on Computer Vision}. Springer, 2022, pp. 709--727.

\bibitem{lian2022scaling}
D.~Lian, D.~Zhou, J.~Feng, and X.~Wang,
\newblock ``Scaling \& shifting your features: A new baseline for efficient model tuning,''
\newblock {\em Advances in Neural Information Processing Systems}, vol. 35, pp. 109--123, 2022.

\bibitem{dong2023efficient}
W.~Dong, D.~Yan, Z.~Lin, and P.~Wang,
\newblock ``Efficient adaptation of large vision transformer via adapter re-composing,''
\newblock {\em arXiv preprint arXiv:2310.06234}, 2023.

\bibitem{zhang2023customized}
K.~Zhang and D.~Liu,
\newblock ``Customized segment anything model for medical image segmentation,''
\newblock {\em arXiv preprint arXiv:2304.13785}, 2023.

\bibitem{lin2023samus}
X.~Lin, Y.~Xiang, L.~Zhang, X.~Yang, Z.~Yan, and L.~Yu,
\newblock ``Samus: Adapting segment anything model for clinically-friendly and generalizable ultrasound image segmentation,''
\newblock {\em arXiv preprint arXiv:2309.06824}, 2023.

\bibitem{wu2023medical}
J.~Wu et~al.,
\newblock ``Medical sam adapter: Adapting segment anything model for medical image segmentation,''
\newblock {\em arXiv preprint arXiv:2304.12620}, 2023.

\bibitem{ma2023segment}
J.~Ma and B.~Wang,
\newblock ``Segment anything in medical images,''
\newblock {\em arXiv preprint arXiv:2304.12306}, 2023.

\bibitem{karimi2019reducing}
D.~Karimi and S.E. Salcudean,
\newblock ``Reducing the hausdorff distance in medical image segmentation with convolutional neural networks,''
\newblock {\em IEEE Transactions on Medical Imaging}, vol. 39, no. 2, pp. 499--513, 2019.

\bibitem{kervadec2019boundary}
H.~Kervadec, J.~Bouchtiba, C.~Desrosiers, E.~Granger, J.~Dolz, and I.B. Ayed,
\newblock ``Boundary loss for highly unbalanced segmentation,''
\newblock in {\em International conference on medical imaging with deep learning}. PMLR, 2019, pp. 285--296.

\bibitem{sun2023boundary}
F.~Sun, Z.~Luo, and S.~Li,
\newblock ``Boundary difference over union loss for medical image segmentation,''
\newblock in {\em International Conference on Medical Image Computing and Computer-Assisted Intervention}. Springer, 2023, pp. 292--301.

\bibitem{hssayeni2020intracranial}
M.D. Hssayeni, M.S. Croock, A.D. Salman, H.F Al-khafaji, Z.A. Yahya, and B.~Ghoraani,
\newblock ``Intracranial hemorrhage segmentation using a deep convolutional model,''
\newblock {\em Data}, vol. 5, no. 1, pp. 14, 2020.

\bibitem{li2023state}
X.~Li et~al.,
\newblock ``The state-of-the-art 3d anisotropic intracranial hemorrhage segmentation on non-contrast head ct: The instance challenge,''
\newblock {\em arXiv preprint arXiv:2301.03281}, 2023.

\bibitem{ronneberger2015u}
O.~Ronneberger, P.~Fischer, and T.~Brox,
\newblock ``U-net: Convolutional networks for biomedical image segmentation,''
\newblock in {\em Medical Image Computing and Computer-Assisted Intervention}. Springer, 2015, pp. 234--241.

\bibitem{oktay2018attention}
O.~Oktay et~al.,
\newblock ``Attention u-net: Learning where to look for the pancreas,''
\newblock {\em arXiv preprint arXiv:1804.03999}, 2018.

\bibitem{zhou2018unet++}
Z.~Zhou, MM~Rahman Siddiquee, N.~Tajbakhsh, and J.~Liang,
\newblock ``Unet++: A nested u-net architecture for medical image segmentation,''
\newblock in {\em Deep Learning in Medical Image Analysis and Multimodal Learning for Clinical Decision Support}. Springer, 2018, pp. 3--11.

\bibitem{chen2021transunet}
J.~Chen et~al.,
\newblock ``Transunet: Transformers make strong encoders for medical image segmentation,''
\newblock {\em arXiv preprint arXiv:2102.04306}, 2021.

\bibitem{zhang2021transfuse}
Y.~Zhang, H.~Liu, and Q.~Hu,
\newblock ``Transfuse: Fusing transformers and cnns for medical image segmentation,''
\newblock in {\em Medical Image Computing and Computer-Assisted Intervention}. Springer, 2021, pp. 14--24.

\bibitem{he2023h2former}
A.~He, K.~Wang, T.~Li, C.~Du, S.~Xia, and H.~Fu,
\newblock ``H2former: An efficient hierarchical hybrid transformer for medical image segmentation,''
\newblock {\em IEEE Transactions on Medical Imaging}, 2023.

\end{thebibliography}

\end{document}